\documentclass[twocolumn,showpacs,preprintnumbers,amsmath,amssymb]{revtex4-1}

\usepackage{graphicx}
\usepackage{dcolumn}
\usepackage{simplewick}
\usepackage{bm}
\usepackage{hyperref}
\usepackage[usenames,dvipsnames]{color}

\def\red{|\!|}

\def\pt{\mathbf{T}^{(1)}}

\def\ptt{\mathbf{T}^{(1)}_2}
\def\ps{\mathbf{S}^{(1)}}
\def\pso{\mathbf{S}^{(1)}_1}
\def\pst{\mathbf{S}^{(1)}_2}
\def\helec{{\mathbf{H}}_{\rm elec}^{\rm NSD}}

\begin{document}

\title{Nuclear spin-independent and dependent parity non-conservation in 
       $^{171}$Yb$^+$ using perturbed relativistic coupled-cluster theory}
\author{B. K. Mani}
\affiliation{Department of Physics, University of South Florida, Tampa,
             Florida 33620, USA}
\begin{abstract}

    We present nuclear spin-independent and dependent parity non-conservation 
    amplitudes for the $[4f^{14}]6s\;^2S_{1/2}-[4f^{14}]5d\;^2D_{3/2}$ 
    transition of $^{171}$Yb$^+$, calculated using perturbed relativistic 
    coupled-cluster theory. As a proxy to estimate theoretical uncertainty
    of these results we calculate the excitation energies, hyperfine structure 
    constants and E1 transition amplitudes for the important low lying 
    states. The PNC results presented in paper shall be useful in the
    propose PNC experiments. 
\end{abstract}
\pacs{31.15.bw, 11.30.Er, 31.15.am}


\maketitle


\section{Introduction}

   The theoretical results of atomic parity non-conservation (PNC) when 
combined with the experimental results is an important probe of
physics beyond the standard model of particle physics \cite{khriplovich-91}. 
There are two sources of PNC in atoms, nuclear spin-independent (NSI) and 
nuclear spin-dependent (NSD). The NSI-PNC is well studied and experimentally 
observed in several atoms. The most precise measurement till date is in the 
case of atomic Cs \cite{wood-97}. The same experiment also indicated a 
signature of NSD-PNC effects. The most dominant source of which is the nuclear 
anapole moment (NAM), a parity odd nuclear electromagnetic moment arising from 
parity violating interaction within the nucleus 
\cite{flambaum-80,flambaum-84,zeldovich-58}. However, there are two other
contributions to NSD-PNC, these are the NSD electron-nucleus $Z$ exchange  
interaction and the combined effect of hyperfine interaction and NSI 
electron-nucleus $Z$ exchange  interaction.

  The parameters describing nucleon-nucleon coupling, effect of NAM is subsumed 
into it, extracted from the Cs PNC experiment do not concur with the nuclear 
data \cite{haxton-01}. This certainly calls for the further investigation 
of the NSD-PNC effects in other atomic systems as well. An example of
an alternative experiment is the proposal to measure the PNC in Ba$^+$ ion, 
suggested by Fortson \cite{fortson-93} and is in progress at 
Seattle \cite{sherman-05,sherman-08}. This experiment could lead to an 
unambiguous observation of NAM in the $6s\;^2S_{1/2}-5d\;^2D_{5/2}$
transition, as the NSI-PNC alone does not contribute to this transition. 
It is important to note that the major difficulty to a clear 
observation of NAM is the large NSI signal, which overwhelms the NSD 
signature. The Ra$^+$ ion has also been suggested and is considered to be an 
important candidate for the PNC measurement \cite{wansbeek-08,versolato-10}.
Apart from Ba$^+$ and Ra$^+$ ions which are one-valence systems the other 
promising candidate for PNC, the NAM in particular, measurement is the atomic
Yb. An enhanced effect of PNC has already been reported 
\cite{tsigutkin-09,tsigutkin-10} in neutral Yb, 
the $6s^2\;^1S_0-6s5d\;^3D_2$ transition, and for further refinement 
of the experiment is in progress at Berkeley. The 
$6s\;^2S_{1/2}-5d\;^2D_{3/2}$ transition in Yb$^+$, has also been suggested
to reveal the NAM signature and is being investigated at Los Alamos
\cite{torgerson-10,das-99}.

The atomic theory results using reliable and accurate many-body methods are 
key to estimate the expected value of PNC transition amplitudes and 
extracting NAM. For the theoretical calculations, the relativistic 
coupled-cluster (RCC) theory \cite{coester-58,coester-60} can be of great 
significance, as it is one of the most reliable many-body theory 
to incorporate electron correlation in atomic calculations.  
The RCC has been used extensively in atomic structure calculations 
\cite{eliav-96,pal-07,sahoo-09,nataraj-08,wansbeek-08,pal-09,porsev-10}
of properties like transition energies, hyperfine structure constants, 
electromagnetic transition amplitudes, intrinsic electric dipole moment and 
PNC in atoms. Apart from atomic physics, it has also been used with great 
success in nuclear \cite{hagen-08}, molecular \cite{isaev-04} and the 
condensed matter \cite{bishop-09} physics.  

In this work, we employ perturbed relativistic coupled-cluster (PRCC) theory
to calculate NSI and NSD-PNC amplitudes of the 
$[4f^{14}]6s\;^2S_{1/2}-[4f^{14}]5d\;^2D_{3/2}$ transition in the case of 
$^{171}$Yb$^+$ ion. This is timely as there are few theoretical results,
Sahoo {\em et al} \cite{sahoo-11} and  Dzuba {\em et al} \cite{dzuba-11} for
NSI-PNC and Dzuba {\em et al} \cite{dzuba-11} and 
Porsev {\em et al} \cite{porsev-12} for NSD-PNC are the previous works. 
The NSI-PNC results from Ref. \cite{sahoo-11} calculated using RCC method 
differ substantially from Ref. \cite{dzuba-11} where the 
correlation-potential-method with sum-over-state approach is employed to 
calculate NSI and NSD-PNC. The NSD-PNC results reported in 
Ref. \cite{porsev-12} are based on RPA and, in general, is in agreement with 
the results reported in Ref. \cite{dzuba-11}.  However, the later is based on
the sum-over-state approach, at the level of PNC matrix elements. The 
PRCC method \cite{chattopadhyay-12,latha-09,mani-11} employed in present work 
is different from the sum-over-states approach. It accounts for the all singly 
and doubly excited intermediate states. There are two sets of the cluster 
amplitudes in the PRCC, and the summation over states in the first order 
time-independent perturbation is incorporated in one set of the cluster 
amplitudes. 

  The paper is organized as follows. In Section. \ref{method}, we provide
a brief description of the theoretical methods. The unperturbed RCC equations 
for close-shell and one-valence systems are given to serve as a easy 
reference. The perturbed RCC is then discussed in detail and PRCC equations 
are derived. The expression for E1PNC using PRCC wave function and some 
leading order diagrams are also discussed. Results from the work and
uncertainty estimates are presented and discussed in 
Section. \ref{results}.


\section{Theoretical methods}
\label{method}

\begin{figure}[h]
\begin{center}
  \includegraphics[width = 7.0 cm]{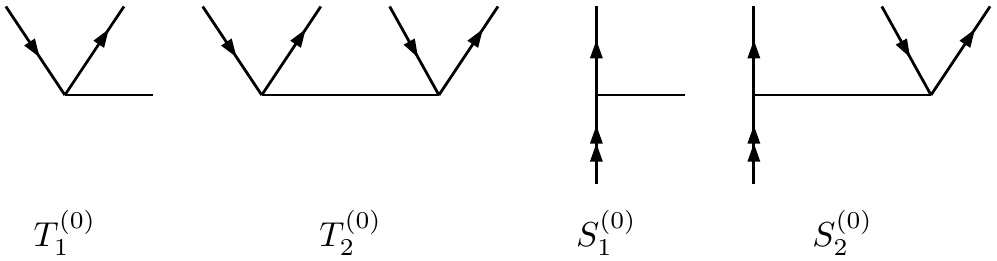}
  \caption{Diagrammatic representation of the single and double excitation
           unperturbed cluster operators in closed shell and one-valence
           sectors.} 
  \label{ts_fig}
\end{center}
\end{figure}

 In absence of PNC interaction the atomic states are of definite parity, and
we consider these as the eigen states of the no-virtual-pair Dirac-Coulomb 
Hamiltonian \cite{sucher-80}
\begin{eqnarray}
  H^{\rm DC}& =& \Lambda _+ \sum_{i=1}^N\left [c\bm{\alpha}_i\cdot \mathbf{p}_i
                 + (\beta_i-1)c^2 - V_N(r_i)\right ]  \nonumber \\
            &  & +\sum_{i<j}\frac{1}{r_{ij}} \Lambda_+,
  \label{dchamil}
\end{eqnarray}
where $\bm{\alpha}_i$ and $\beta$ are the Dirac matrices, $\mathbf{p}$ is the
linear momentum, $V_N(r)$ is the nuclear Coulomb potential and the last term
is the electron-electron Coulomb interactions. The operator $\Lambda_+$
projects on the positive energy eigenstates to avoid the negative energy
continuum solutions. The Hamiltonian $H^{\rm DC}$ satisfies the
eigen value equation
\begin{equation}
  H^{\rm DC}|\Psi_v \rangle = E_v |\Psi_v \rangle,
  \label{hdc_eqn}
\end{equation}
where $|\Psi_v \rangle$ is the exact atomic state of the one-valence system 
and $E_v$ is the corresponding energy. Here after, for compact notation, we
use $H$ to represent $H^{\rm DC}$. In the present work, we use 
RCC theory with the single and doubles (CCSD) excitation approximation to solve 
Eq. (\ref{hdc_eqn}). In RCC, $|\Psi_v \rangle$ is expressed in terms of 
the closed-shell and one-valence cluster operators, $T^{(0)}$ and $S^{(0)}$ 
respectively, as
\begin{equation}
 |\Psi_v\rangle = e^{T^{(0)}} \left [  1 + S^{(0)} \right ] |\Phi_v\rangle,
 \label{psi_unptrb}
\end{equation}
where superscript $(0)$ represents the unperturbed RCC operators. The  
one-valence Dirac-Fock (DF) reference state $|\Phi_v\rangle$ is 
obtained by adding an electron to the closed-shell reference state,
$|\Phi_v \rangle = a^\dagger_v|\Phi_0\rangle$. 
In the CCSD approximation, $T^{(0)} = T^{(0)}_1 + T^{(0)}_2$ 
and $S^{(0)} = S^{(0)}_1 + S^{(0)}_2$. Using the second quantized 
representation
\begin{subequations}
\begin{eqnarray}
  T_1 &=& \sum_{a, p}t_a^p a_p^{\dagger}a_a, \text{ and }
  T_2 = \frac{1}{2!}\sum_{a, b, p, q}t_{ab}^{pq}
  a_p^{\dagger}a_q^{\dagger}a_ba_a, \\ 
  S_1 &=& \sum_{p}s_v^p a_p^{\dagger}a_v, \text{ and }
  S_2 = \sum_{a, p, q}s_{va}^{pq}
  a_p^{\dagger}a_q^{\dagger}a_aa_v.
\end{eqnarray}
\end{subequations}
Here, $t_{\cdots}^{\cdots}$ and $s_{\cdots}^{\cdots}$ are the cluster
amplitudes. The indexes $abc\ldots$ ($pqr\ldots$) represent core (virtual)
states and $vwx\ldots$ represent valence states. The operators $T_1$ ($S_1$ )
and $T_2$ ($S_2$) give single and double replacements after operating on the
closed(open)-shell reference states. The diagrammatic representation of
these operators are shown in Fig. \ref{ts_fig}.

%
%

   The open-shell cluster operators are then the solutions of nonlinear 
equations \cite{mani-10}
\begin{subequations}
\label{s0_eqn}
\begin{eqnarray}
  \langle \Phi_v^p|\bar H_N \! +\! \{\contraction[0.5ex]
  {\bar}{H}{_N}{S} \bar H_N S^{(0)}\} |\Phi_v\rangle
  &=&E_v^{\rm att}\langle\Phi_v^p|S^{(0)}_1|\Phi_v\rangle ,
  \label{s01_eqn}     \\
  \langle \Phi_{va}^{pq}|\bar H_N +\{\contraction[0.5ex]
  {\bar}{H}{_N}{S}\bar H_N S^{(0)}\} |\Phi_v\rangle
  &=& E_v^{\rm att}\langle\Phi_{va}^{pq}|S^{(0)}_2|\Phi_v\rangle,
  \label{s02_eqn}
\end{eqnarray}
\end{subequations}
where $\bar H_{\rm N}=e^{-T^{(0)}}H_{\rm N}e^{T^{(0)}} $ is the
similarity transformed Hamiltonian, 
$H_{\rm N} = H -\langle\Phi_0|H|\Phi_0\rangle$ is the normal ordered 
Hamiltonian  and $E_v^{\rm att}$ is the attachment energy of the
valence electron. The operators $T^{(0)}$ are the solutions of a similar 
set of nonlinear coupled equations
\begin{subequations}
\label{t0_eqn}
\begin{eqnarray}
  \langle\Phi^p_a|\bar H_{\rm N}|\Phi_0\rangle = 0, 
     \label{t01_eqn}                        \\
  \langle\Phi^{pq}_{ab}|\bar H_{\rm N}|\Phi_0\rangle = 0.
     \label{t02_eqn} 
\end{eqnarray}
\end{subequations}
The details on the derivation of these equations are given in 
our previous work \cite{mani-09}.

     In presence of PNC interaction atomic states mix with the opposite 
parity states and the total atomic Hamiltonian is 
\begin{equation}
    H_{\rm a} = H^{\rm DC} + \lambda H_{\rm PNC}.
    \label{total_H}
\end{equation}
Here, $\lambda$ is the perturbation parameter and $H_{\rm PNC}$
represents the any general PNC interaction Hamiltonian. It has two components,
the NSI and NSD interaction. These are 
\begin{subequations}
\begin{eqnarray}
   H_{\rm PNC}^{\rm NSI}&=&\frac{G_{\rm F}Q_W}{2 \sqrt{2}} \sum_i
   \gamma_5\rho_{\rm{N}}(r_i), 
  \label{hpnc_nsi}                \\
   H_{\rm PNC}^{\rm NSD}&=&\frac{G_{\rm F}\mu'_W}{\sqrt{2} I}\sum_i
   \bm{\alpha}_i\cdot \mathbf{I}\rho_{\rm{N}}(r),
  \label{hpnc_nsd}
\end{eqnarray}
\end{subequations}
where, $G_F (=2.22\times10^{-14} a.u.)$ is the Fermi coupling 
constant, $Q_W$ and $\mu'_W$ are respectively the weak nuclear charge and 
the weak nuclear moment of the nucleus expressed in terms of neutron and 
proton numbers, $\bm{\alpha}$ and $\gamma_5$ are the Dirac matrices, 
$\rho_{\rm N}(r)$ is the normalized nuclear density and $I$ is the 
nuclear spin. Compared to the NSI-PNC, the NSD-PNC require two important
considerations because of the nuclear spin operator $\mathbf{I}$. First, the 
cluster operators in the electron space are rank one operators, and second, 
the atomic states in the one-valence sector are eigenstates of total angular 
momentum $\mathbf{F} = \mathbf{I} + \mathbf{J}$.

%
%
\begin{figure}[h]
\begin{center}
  \includegraphics[width = 7.0 cm]{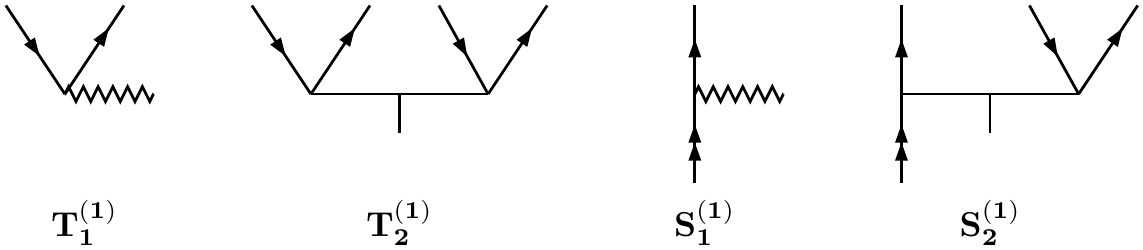}
  \caption{Diagrammatic representation of the single and double excitation
           NSD-perturbed cluster operators in closed-shell and one-valence
           sectors. The extra line in the $T_2^{(1)} $ and $S_2^{(1)}$ is to
           indicate the multipole structure of the operators.}
  \label{pts_nsd_fig}
\end{center}
\end{figure}

     Similar to the unperturbed eigen value equation, Eq. (\ref{hdc_eqn}),
we may write the perturbed eigenvalue equation, satisfied by the total
atomic Hamiltonian, as
\begin{equation}
  H_{\rm a} |\widetilde{\Psi}_v \rangle = 
  \widetilde{E}_v |\widetilde{\Psi}_v \rangle,
  \label{ht_eqn}
\end{equation}
where $|\widetilde{\Psi}_v \rangle$ is the perturbed atomic state
and $\widetilde{E}_v$ is the corresponding energy. To the first-order 
in $\lambda $, 
$|\widetilde{\Psi}_v \rangle = |\Psi_v\rangle + 
\lambda |\bar{\Psi}^{1}_v\rangle$ and 
$\widetilde{E}_v = E_v + \lambda E^{1}_v$, where the bar in
$|\bar{\Psi}^{1}_v\rangle$ denotes it's parity is opposite to $|\Psi_v\rangle$.
From here on, to derive the PRCC equations we consider the NSD-PNC interaction
Hamiltonian. Using Eq. (\ref{hpnc_nsd}), we can rewrite Eq. (\ref{ht_eqn}) as
\begin{equation}
  \left ( H^{\rm DC} + \lambda \helec\cdot\mathbf{I} \right) | 
  \widetilde{\Psi}_v \rangle = E_v| \widetilde{\Psi}_v \rangle.
  \label{ht_elc_eqn}
\end{equation}
Here, $H_{\rm elc}^{\rm NSD} =( G_{\rm F}\mu'_W/\sqrt{2})\sum_i
\alpha_i\rho_{\rm{N}}(r)$ is the electronic part of $H_{\rm PNC}^{\rm NSD}$. 
While writing above equation we have used 
$E^1_v = \langle \Psi_v|H_{\rm PNC}^{\rm NSD}|\Psi_v\rangle = 0$,
as $H_{\rm PNC}^{\rm NSD}$ is an odd parity operator it connects
opposite parity states only. In the PRCC theory, the perturbed wave function
is expressed as 
\begin{equation}
  | \widetilde{\Psi}_v \rangle = e^{T^{(0)}}\left[ 1 
    + \lambda \pt \cdot\mathbf{I} \right] \left[ 1
    + S^{(0)} +\lambda \ps \cdot\mathbf{I} \right] |\Phi_v \rangle,
  \label{psi_ptrb}
\end{equation}
where $\pt$ and $\ps$ are  the closed-shell and one-valence PRCC operators, 
respectively. The superscript $(1)$ is used to indicate the perturbation. 
The diagrammatic representation of these cluster operators are shown in 
Fig. \ref{pts_nsd_fig}.  

      Using Eq. (\ref{psi_ptrb}) in Eq. (\ref{ht_elc_eqn}), we can rewrite 
the eigenvalue equation as
\begin{eqnarray}
  &&\left( H + \lambda\helec \cdot\mathbf{I} \right)
  e^{T^{(0)}} \left[ 1 + \lambda \pt \cdot\mathbf{I} \right]
  \left[ 1 + S^{(0)} \right . \nonumber \\
 && \left . + \lambda\ps\cdot\mathbf{I} \right] |\Phi_v \rangle 
  = E_v e^{T^{(0)}} \left[ 1 + \lambda \pt\cdot\mathbf{I} \right]
    \left[ 1 + S^{(0)} \right .  \nonumber \\
 && \left. + \lambda \ps\cdot\mathbf{I} \right] |\Phi_v \rangle.
\end{eqnarray}
To derive the PRCC equations, we project the above equation on $e^{-T^{(0)}}$ 
and retain the terms linear in $\lambda$. In addition, for further 
simplification, we use normal-ordered form of the Hamiltonian
$H_{\rm N} = H - \langle\Phi_v|H|\Phi_v\rangle$. After these sequence of 
operations, the eigenvalue equation is modified to
\begin{eqnarray}
  \left[ \bar H_{\rm N}\ps  + \bar H_{\rm N}\pt ( 1 + S^{(0)} ) +
    \bar {\mathbf{H}}_{\rm elec}^{\rm NSD} ( 1 + S^{(0)} ) \right]
    |\Phi_v \rangle      \nonumber \\
  =\left[ \Delta E_v \ps + \Delta E_v \pt ( 1 + S^{(0)} ) \right]|\Phi_v 
     \rangle,
  \label{deltae1v}
\end{eqnarray}
where $\Delta E_v = E_v - \langle\Phi_v|H|\Phi_v\rangle$, is the correlation
energy of the one-valence system. Like $\bar{H}_{\rm N}$ introduced earlier, 
$\bar {\mathbf{H}}_{\rm elec}^{\rm NSD} 
=e^{-T^{(0)}}H_{\rm elc}^{\rm PNC}e^{T^{(0)}}$ is the similarity 
transformed NSD-PNC interaction  Hamiltonian in the electronic space.
The PRCC equations of $\ps$ can now be derived by projecting 
Eq. (\ref{deltae1v}) with the excited determinants $\langle\Phi^p_v|$ and 
$\langle\Phi^{pq}_{va}|$ as
\begin{widetext}
\begin{subequations}
\begin{eqnarray}
   \langle \Phi^p_v |\{ \contraction[0.5ex]{}{H}{_{\rm N}}{S}\bar{H}_{\rm N}
     \mathbf{S}^{(1)} \} + \{ \contraction{}{H}{_{\rm N}}{S}\bar{H}_{\rm N}
     \mathbf{T}^{(1)} \} + \{ \contraction[0.5ex]{}{H}{_{\rm N}}{T}
     \contraction[0.8ex]{}{V}{_{\rm N}T^{(1)}}{S}\bar{H}_{\rm N}
     \mathbf{T}^{(1)}S^{(0)}\} + \bar{\mathbf{H}}_{\rm elec}^{\rm NSD} 
    + \{ \contraction[0.5ex]{}{H}{_{\rm elec}^{\rm NSD}}{S}
     \bar{\mathbf{H}}_{\rm elec}^{\rm NSD}{S}^{(0)} \}|\Phi_v \rangle &=&
  E_v^{\rm att} \langle \Phi^p_v | \pso|\Phi_v \rangle, \\
  \langle \Phi^{pq}_{vb}|\{ \contraction[0.5ex]{}{H}{_{\rm N}}{S}\bar{H}_{\rm N}
     \mathbf{S}^{(1)}\}+\{ \contraction[0.5ex]{}{H}{_{\rm N}}{S}\bar{H}_{\rm N}
     \mathbf{T}^{(1)} \} + \{ \contraction[0.5ex]{}{H}{_{\rm N}}{T}
     \contraction[0.8ex]{}{V}{_{\rm N}T^{(1)}}{S}\bar{H}_{\rm N}
     \mathbf{T}^{(1)}S^{(0)}\} + \bar{\mathbf{H}}_{\rm elec}^{\rm NSD} 
    + \{ \contraction[0.5ex]{}{H}{_{\rm elec}^{\rm NSD}}{S}
     \bar{\mathbf{H}}_{\rm elec}^{\rm NSD}{S}^{(0)} \}|\Phi_v \rangle &=&
   E_v^{\rm att} \langle \Phi^{pq}_{vb} | \pst|\Phi_v \rangle.
  \label{ccsptrb1v2}
\end{eqnarray}
\label{prcc_eqn}
\end{subequations}
\end{widetext}
While deriving the equations we have used the relations 
$ \langle \Phi^p_v | \pt |\Phi_v \rangle = 0$ and
$\langle \Phi^p_v| \pt S| \Phi_v \rangle = 0$. These follows as $\pt$  is an
operator of closed-shell sector, it does not contribute to the PRCC equation of
$\pso$ and $\pst$. The closed-shell operators $\pt$ are the solutions
of the similar set of coupled equations \cite{mani-09}
\begin{subequations}
\label{pcceq}
\begin{eqnarray}
  \langle \Phi^p_a |\{ \contraction{}{H}{_{\rm N}}{T}
     \bar{\mathbf{H}}_{\rm N}\mathbf{T}^{(1)} \} |\Phi_0\rangle &=&
  -\langle \Phi^p_a | \bar {\mathbf{H}}_{\rm elec}^{\rm NSD}
      - \Delta E_0 \pt |\Phi_0 \rangle, \;\;\;\;\;\;\;\;
  \label{pcceq1}                         \\
  \langle \Phi^{pq}_{ab} | \{\contraction{}{H}{_{\rm N}}{T}
  \bar{\mathbf{H}}_{\rm N}\mathbf{T}^{(1)} \} |\Phi_0 \rangle &=&
  -\langle \Phi^{pq}_{ab} | \bar {\mathbf{H}}_{\rm elec}^{\rm NSD} 
   - \Delta E_0 \pt  |\Phi_0 \rangle .
 \label{pcceq2}
\end{eqnarray}
\end{subequations}  
These equations can be derived from the closed-shell perturbed eigenvalue
equation. We can also derive a similar set of PRCC equations for the NSI-PNC
interaction Hamiltonian. One major difference is, the cluster operators are
rank zero operators.  

\begin{figure}[h]
\begin{center}
  \includegraphics[width = 8.0cm]{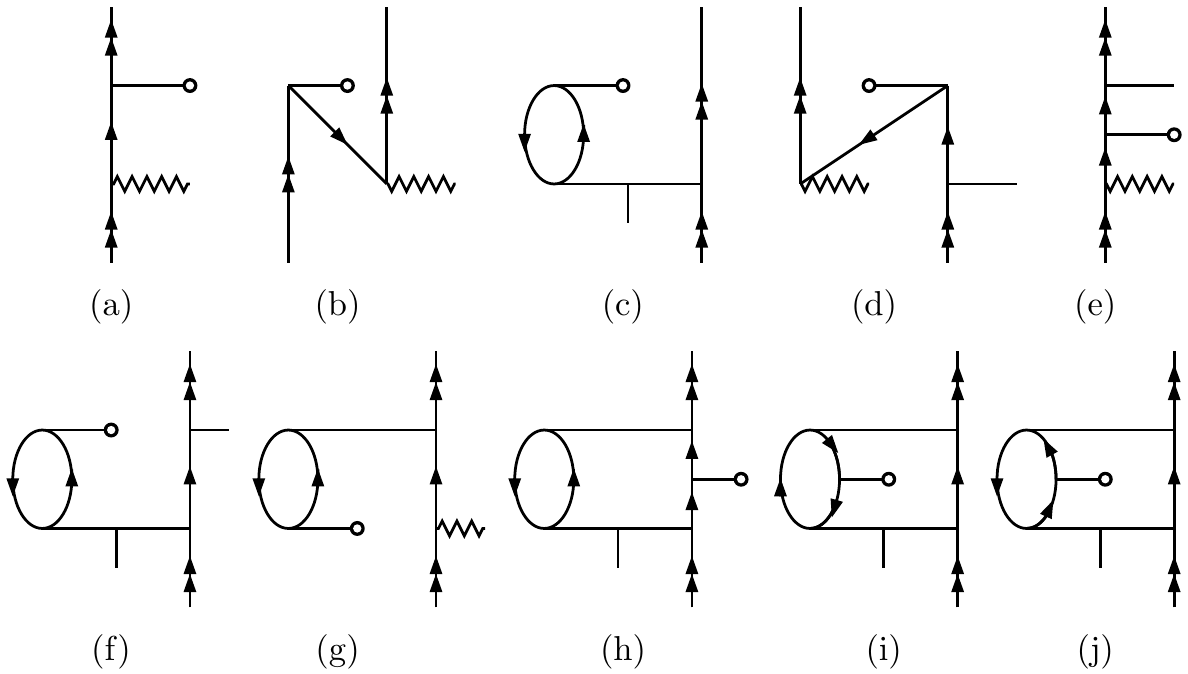}
  \caption{Some of the leading order PRCC diagrams which contribute to the
           $E1_{\rm elec}^{\rm PNC}$ of one-valence atoms.}
  \label{e1pnc_cc_fig}
\end{center}
\end{figure}

     After solving the RCC and PRCC equations, we can use the atomic states 
for the properties calculations. The RCC expressions and the diagrams 
contributing to the hyperfine structure (HFS) constants and the E1 transition 
amplitudes are derived and discussed in our previous work \cite{mani-10}. In 
the present work, we use the same expressions and diagrams to compute
HFS constants and E1 transition amplitudes. The PNC induced electric dipole
transition amplitude, using PRCC wave function, is
\begin{equation}
  {\rm E1PNC} = \langle \widetilde{\Psi}_w\red \mathbf{D} \red 
                 \widetilde{\Psi}_v \rangle,
\end{equation}
where $\mathbf{D}$ is the dipole operator. This expression, unlike the 
conventional sum-over-sates approach, implicitly account for all the possible
intermediate states. From Eq. (\ref{psi_ptrb}), for the NSD-PNC interaction,
the transition amplitude is
\begin{eqnarray}
   E1_{\rm PNC}^{\rm NSD} && =  \langle \Phi_w \red {e^{T^{(0)}}}^\dagger 
       \left[ 1 + \lambda \pt\cdot\mathbf{I}  \right]^\dagger \left[ 1 
       + S^{(0)} + \lambda \ps\cdot\mathbf{I} \right]^\dagger 
                          \nonumber \\
       && \mathbf{D} e^{T^{(0)}} \left[ 1 + \lambda \pt\cdot\mathbf{I} \right]
       \left[ 1 + S^{(0)} + \lambda \ps\cdot\mathbf{I}\right] \red 
       \Phi_v \rangle.
\end{eqnarray}
Consider terms linear in $\lambda$ and retain only those up to second order 
in cluster amplitude. Define the electronic component as
$E1_{\rm elec}^{\rm NSD} $, corresponding to the $H_{\rm elec}^{\rm NSD}$, 
it is then given as 
\begin{eqnarray}
   E1_{\rm elec}^{\rm NSD}&  \approx & \langle\Phi_w\red 
               \mathbf{D} \pt + {T^{(0)}}^\dagger \mathbf{D} \pt + 
               {\pt}^\dagger \mathbf{D} T^{(0)}  
                                                 \nonumber \\
            && +{\pt}^\dagger \mathbf{D} + \mathbf{D} \pt S^{(0)} 
               + {\pt}^\dagger {S^{(0)}}^\dagger \mathbf{D} 
                                                 \nonumber \\
            && +{S^{(0)}}^\dagger \mathbf{D} \pt
               + {\pt}^\dagger \mathbf{D} S^{(0)} + \mathbf{D} \ps
               +{\ps}^\dagger \mathbf{D}                      \nonumber \\
            && + {S^{(0)}}^\dagger \mathbf{D} 
               \ps + {\ps}^\dagger \mathbf{D} S^{(0)} \red\Phi_v\rangle.
  \label{e1pnccc1v}
\end{eqnarray}
To calculate ${\rm E1PNC}$, we use diagrammatic analysis to identify the 
Goldstone diagrams from these terms. However, we exclude the structural 
radiation diagrams, arising from the terms involving two-body cluster 
operators, for example,  $\ptt\mathbf{D}T^{(0)}_2$. The selected diagrams 
from the leading order and next to leading order terms are shown in the 
Fig. \ref{e1pnc_cc_fig}.

%
%
%
%

 
 \section{Results and discussions}
\label{results}


\subsection{Single-particle basis functions}

  For all the calculations we use Gaussian type orbitals (GTOs) or single
particle wave functions with  $V^{N-2}$ potential. As mentioned earlier, to
incorporate the relativistic effects we use the Dirac-Coulomb atomic
Hamiltonian. For the nuclear potential we consider the  finite size Fermi 
density distribution
\begin{equation}
  \rho_{\rm nuc}(r) = \frac{\rho_0}{1 + e^{(r-c)/a} },
\end{equation}
here, $a = t 4\ln 3$. The parameter $c$ is the half-charge radius, that is
$\rho_{\rm nuc}(c)=\rho_0/2$ and $t$ is the skin thickness. The orbitals
are of the form
\begin{equation}
  \psi_{n\kappa m}(\bm{r})=\frac{1}{r}
  \left(\begin{array}{r}
            P_{n\kappa}(r)\chi_{\kappa m}(\bm{r}/r)\\
           iQ_{n\kappa}(r)\chi_{-\kappa m}(\bm{r}/r)
       \end{array}\right),
  \label{spin-orbital}
\end{equation}
where $P_{n\kappa}(r)$ and $Q_{n\kappa}(r)$ are the large and small component
radial wave functions, $\kappa$ is the relativistic total angular momentum
quantum number and $\chi_{\kappa m}(\bm{r}/r)$ are the spinor-spherical
harmonics. The radial components are then defined as linear combination of 
Gaussian type functions \cite{mohanty-89,chaudhuri-99} 
\begin{eqnarray}
   P_{n\kappa}(r) = \sum_p C^L_{\kappa p} g^L_{\kappa p}(r),  \nonumber \\
   Q_{n\kappa}(r) = \sum_p C^S_{\kappa p} g^S_{\kappa p}(r).
\end{eqnarray}
The index $p=1, 2, \ldots, m$, where $m$ is the number of basis functions and
$C_{\kappa p}^{\cdots}$ are the coefficients of linear combination. For large 
component we choose
\begin{equation}
  g^L_{\kappa p}(r) = C^L_{m_{\kappa i}} r^{n_\kappa} e^{-\alpha_p r^2},
\end{equation}
where $n_\kappa$ is an integer and $C^L_{m_{\kappa i}}$ is the normalization 
constant. The small component are derived from the large components using 
kinetic balance condition. The $\alpha_p$ follow the general relation
\begin{equation}
  \alpha_p = \alpha_0 \beta^{p-1}.
  \label{param_gto}
\end{equation}
The parameters, $\alpha_0$ and $\beta$, are optimized such that the single
particle energies of the core and valence orbitals are in good agreement with
the numerical results, obtained from GRASP92 \cite{parpia-96}. In 
Table. \ref{grasp_e_tab}, we compare the energy of the valence orbitals 
from the GTO with the GRASP92 data. 
%
%
\begin{table}[h]
\begin{center}
\caption{The valence orbital and SCF energies of Gaussian type orbitals (GTO) 
         are compared with the GRASP92 data.}
\begin{ruledtabular}
\begin{tabular}{ccc}
  Orbitals  &  GTO   & GRASP92 \\
\hline
  $6s\;^2S_{1/2}$ & $-0.413668$ & $-0.413665$ \\
  $6p\;^2P_{1/2}$ & $-0.301112$ & $-0.301113$ \\
  $6p\;^2P_{3/2}$ & $-0.288305$ & $-0.288307$ \\
  $5d\;^2D_{3/2}$ & $-0.303070$ & $-0.303071$ \\
  $5d\;^2D_{5/2}$ & $-0.300885$ & $-0.300886$ \\
  $E_{\rm SCF}$ & $-14067.0622$ & $-14067.0676$ 
\end{tabular}
\end{ruledtabular}
\label{grasp_e_tab}
\end{center}
\end{table}
%
%


\subsection{Excitation energies, hyperfine structure constants 
            and E1 transition amplitudes}

\begin{figure}[h]
\begin{center}
  \includegraphics[width = 6.0cm, angle = -90]{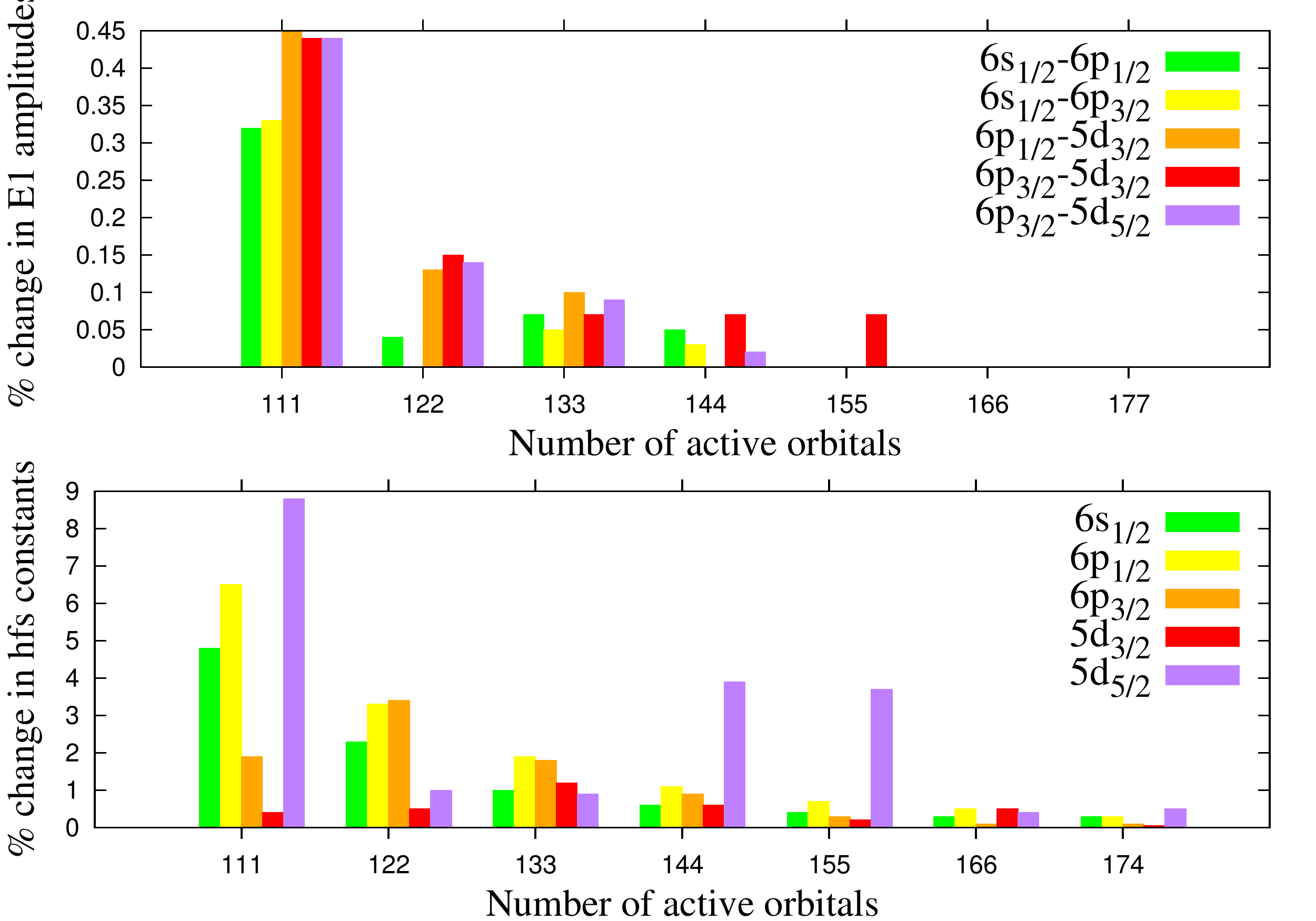}
  \caption{The convergence (\% change) of the hyperfine structure constants 
           and the E1 transition amplitudes with respect to the number of
           active orbitals.} 
  \label{hfs_fig}
\end{center}
\end{figure}

The excitation energies, hyperfine structure constants and the E1 transition 
amplitudes from our calculations are listed in the 
Tables. \ref{ee_tab}, \ref{hfs_tab} and \ref{e1_tab}, respectively. 
These results are obtained using a fairly large basis of 
$177$ active GTOs, it consists of  $19$, $17$, $17$, $17$, $15$ and $13$ 
orbitals in the $s$, $p$, $d$, $f$, $g$ and $h$ symmetries, respectively. 
To arrive at this basis set we start with a moderate size of $100$ 
active orbitals with the combination $12s$, $10p$, $10d$, $10f$, $8g$ and 
$6h$. And perform seven sets of calculations by adding one orbital to each
symmetry in every successive sets. The \% change in HFS
constants and E1 transition amplitudes with respect to number of
active orbitals are shown in Fig. \ref{hfs_fig}. As we see in the
figure, the E1 transition amplitudes converge and there is no observable
change in the amplitudes after $155$. On the other hand, for HFS constants 
we observe a slower convergence pattern. It is evident from the figure that
the  HFS results are close to convergence. The maximum uncertainty is about 
0.5\%, in the case of $5d\;^2D_{5/2}$, but it is smaller for the states 
$6s\;^2S_{1/2}$, $6p\;^2P_{1/2}$, $6p\;^2P_{3/2}$ and $5d\;^2D_{3/2}$, the
uncertainties are 0.3\%, 0.3\%, 0.1\% and 0.05\%, respectively. 
%
%
%
%
\begin{table}[h]                                                                
\begin{center}                                                               
\caption{Excitation energy for some of the low lying excitations in
         $^{171}$Yb$^+$. The values are in cm$^{-1}$.}                                    
 \begin{ruledtabular}                                                         
\begin{tabular}{cccc}                                                       
Level & This work & Other works & Exp.{Ref\cite{nist}.} \\                  
\hline                                                                        
$5d_{3/2}$ & $23983$ & $23926^{\rm a}$   & $22961$ \\
           &         & $21238^{\rm b}$   &           \\
           &         & $22711^{\rm c}$   &           \\
           &         & $22820^{\rm d}$   &           \\
$5d_{5/2}$ & $25576$ & $22449^{\rm b}$   & $24333$ \\
           &         & $24178^{\rm c}$   &           \\
           &         & $24261^{\rm d}$   &           \\
$6p_{1/2}$ & $27985$ & $28749^{\rm a}$   & $27062$ \\
           &         & $28048^{\rm b}$   &           \\
           &         & $27945^{\rm c}$   &           \\
           &         & $27945^{\rm d}$   &     \\     
           &         & $28109(1000)^{\rm e}$   &     \\
$6p_{3/2}$ & $31757$ & $32376^{\rm a}$   & $30392$ \\
           &         & $31411^{\rm b}$   &           \\       
           &         & $31403^{\rm c}$   &           \\       
           &         & $31481^{\rm d}$   &           \\       
           &         & $31604(800)^{\rm e}$   &      \\       
\end{tabular}                                                                
\end{ruledtabular}                                                           
\begin{tabbing}                                                              
  $^{\rm a}$ Reference\cite{dzuba-11}. \\
  $^{\rm b}$ Reference\cite{safronova-09}.\\
  $^{\rm c}$ Reference\cite{porsev-12}-MBPT + corrections.\\
  $^{\rm d}$ Reference\cite{porsev-12}-All-order. \\
  $^{\rm e}$ Reference\cite{sahoo-11}.                                       
\end{tabbing}                                                                
\label{ee_tab}                                                              
\end{center}                                                                 
\end{table}                                                                 

 The excitation energies from our calculations are listed in 
Table. \ref{ee_tab}. As described in Sec. \ref{method}, these are calculated 
using RCC with the CCSD approximation. Except for the $6p\;^2P_{3/2}$ 
excitation energy, our results are better or on par with the previous 
theoretical results when compared with the experimental data. The all-order 
results reported in Ref. \cite{porsev-12} are closer to the experimental
data than the other theoretical results, including the present work.
For the $6p\;^2P_{3/2}$, our result is in close to the RCC result of 
Sahoo and collaborators \cite{sahoo-11}. Among the other three results for this 
level, our result is closer to the Ref. \cite{porsev-12}.
The dominant excitations which contribute to the denominator of the
NSI-PNC matrix are $6s\;^2S_{1/2}-6p\;^2P_{1/2}$ and 
$6p\;^2P_{1/2}-5d\;^2D_{3/2}$.  However for the NSD-PNC, these are 
$6s\;^2S_{1/2}-6p\;^2P_{1/2}$, $6p\;^2P_{1/2}-5d\;^2D_{3/2}$,
$6s\;^2S_{1/2}-6p\;^2P_{3/2}$ and $6p\;^2P_{3/2}-5d\;^2D_{3/2}$. The accuracy 
achieved for these in the present work are 3.4\%, 4.5\%, 2.4\% and 4.6\%,
respectively. We have incorporated these errors in the total uncertainty
estimates for the PNC results. 
%
%
\begin{table}[h]                                                                 
\begin{center}                                                                
\caption{Magnetic dipole hyperfine structure constants of $^{171}$Yb$^+$ in
         the unit MHz.}
\begin{ruledtabular}                                                          
\begin{tabular}{ccccc} 
State & This work & Other works & Exp            \\  
\hline                                                             
$6s_{1/2}$ & $13488.314$& $13217^{\rm a},13172^{\rm b},$
                        & $12645(2)^{\rm e}$ \\       
           &            & $13091^{\rm c},13332(1000)^{\rm d},$ \\
           &            & $12730(2)^{\rm e}$ & \\

$6p_{1/2}$ & $2348.036$ & $2533^{\rm a},2350^{\rm b},$
                        & $2104.9(1.3)^{\rm e}$ \\       
           &            & $2371^{\rm c},2516(400)^{\rm d},$ \\ 
           &            & $2317^{\rm e}$ & \\

$6p_{3/2}$ & $313.522$  & $388^{\rm a},311.5^{\rm b},330^{\rm c},$ 
                        & $877(20)^{\rm f}$ \\       
           &            & $322(20)^{\rm d},391^{\rm e}$& \\

$5d_{3/2}$ & $421.131$ & $291^{\rm a},489^{\rm c},
                         447(20)^{\rm d},400.5^{\rm g},$ 
                        & $430(43)^{\rm h}$ \\

$5d_{5/2}$ & $-68.567$ & $-96^{\rm c},-48(15)^{\rm d},-12.6^{\rm g}$ 
                                    & $-63.6(7)^{\rm i}$\\       
\end{tabular}    
\end{ruledtabular}
\begin{flushleft}
$^{\rm a}$ Reference\cite{dzuba-11},
$^{\rm b}$ Reference\cite{safronova-09},
$^{\rm c}$ Reference\cite{porsev-12},\\
$^{\rm d}$ Reference\cite{sahoo-11}, 
$^{\rm e}$ Reference\cite{martensson-94}, 
$^{\rm f}$ Reference\cite{berends-92},\\
$^{\rm g}$ Reference\cite{itano-06},                                    
$^{\rm h}$ Reference\cite{engelke-96},                                    
$^{\rm i}$ Reference\cite{roberts-99}.                                    
\end{flushleft} 
  \label{hfs_tab}
\end{center} 
\end{table}

In the Table. \ref{hfs_tab} we present, and compare HFS constants obtained
from the present calculations with the other theoretical and experimental 
results. As evident from the table, our results for the $6p\;^2P_{1/2}$, 
$5d\;^2D_{3/2}$ and $5d\;^2D_{5/2}$ states are in better agreement with the
experimental data than the other theoretical results. However, for the 
$6s\;^2S_{1/2}$ state, like the other theoretical results, our result is also 
larger than the experimental result. Among the theoretical results, our 
result for this state is closer to the Ref. \cite{sahoo-11}. The reason for
this is the method employed and the type of the orbitals used in the two 
works are similar. For the $6p\;^2P_{3/2}$ state there is a large discrepancy 
between the theoretical results and experimental data. However, it must be
emphasized that the experimental data is from a relatively old measurement. 
Our result lies between the third-order MBPT results of 
Ref. \cite{safronova-09} and the RCC result of Ref. \cite{sahoo-11}.

 The impact of the electron correlation effects is discernible in the 
Table. \ref{hfscompo_tab}, where we list the contributions from various 
RCC terms. The RCC terms in the table are based on the expression in our 
previous work \cite{mani-10}. The contribution listed as ``Other'' correspond 
to the terms $S_2^\dagger H_{\rm hfs} T + {\rm c.c.}$ and 
$S_2^\dagger H_{\rm hfs}TS_1+{\rm c.c.}$. 
As expected, the dominant contribution is from the DF term. 
It contributes approximately about 72\%, 66\%, 58\%, 69\% and 110\% for
the states $6s\;^2S_{1/2}$, $6p\;^2P_{1/2}$, $6p\;^2P_{3/2}$, 
$5d\;^2D_{3/2}$ and $5d\;^2D_{5/2}$, respectively.
Our DF value $9716.7$ and $1548.2$ for states $6s\;^2S_{1/2}$ and 
$6p\;^2P_{1/2}$, respectively are on the higher side of the values, 
$9577$ and $1542$, reported by Safronova and collaborators in their recent 
work \cite{porsev-12}. On the other hand, for the $6p\;^2P_{3/2}$, 
$5d\;^2D_{3/2}$ and $5d\;^2D_{5/2}$ states our results of $182.5$, $289.7$ 
and $110.2$, show a close match with the values $183$, $290$ and $111$ from 
Ref. \cite{porsev-12}. The next two leading order contributions are from
the terms $S_1^\dagger \tilde{H}_{\rm hfs} +{\rm c.c.}$ and 
$S_2^\dagger \tilde{H}_{\rm hfs} +{\rm c.c.}$. Unlike other states, 
$5d\;^2D_{5/2}$ shows a different correlation pattern and contribution from 
$S_2^\dagger \tilde{H}_{\rm hfs} +{\rm c.c.}$ is about -321\% of the total 
value. However $S_1^\dagger \tilde{H}_{\rm hfs} +{\rm c.c.}$ contributes 
44\%  of the total value. Despite the large cancellations our total 
result compares well with the experiment.

\begin{table}[h]
\begin{center} 
\caption{The electric dipole transition amplitudes of $^{171}$Yb$^+$.} 
\begin{ruledtabular}
\begin{tabular}{cccc}
Transition & This work  &  Other works & Exp. \\
\hline                                                                        
$6p_{1/2}\longleftarrow 6s_{1/2}$ & $2.748$  & $2.72^{\rm a},2.73^{\rm b}$,
                                          & $2.47(3)^{\rm f}$ \\
                               &          & $2.75^{\rm c},2.64^{\rm d},
                                             2.72(1)^{\rm e}$ \\ 

$6p_{3/2}\longleftarrow 6s_{1/2}$ & $3.901$  & $3.84^{\rm a},3.84^{\rm b}$,
                                          & $3.36(3)^{\rm g}$  \\
                               &          & $3.83^{\rm c},3.71^{\rm d},
                                             3.83(1)^{\rm e}$ \\

$5d_{3/2}\longleftarrow 6p_{1/2}$ & $3.138$  & $3.09^{\rm a},3.78^{\rm b}$,
                                          & $2.97(4)^{\rm f}$ \\
                               &          & $3.06^{\rm c},2.98^{\rm d},
                                             3.06(2)^{\rm e}$ \\

$5d_{3/2}\longleftarrow 6p_{3/2}$ & $1.369$  & $1.36^{\rm a},1.55^{\rm b}$
                                          & $-$  \\
                               &          & $1.35^{\rm c},1.32^{\rm d},
                                             1.35(2)^{\rm e}$ \\

$5d_{5/2}\longleftarrow 6p_{3/2}$ & $4.307$ & $4.77^{\rm b},4.23^{\rm c}$ 
                                            & $-$ \\
                               &            & $4.23(3)^{\rm e}$
\end{tabular}
\end{ruledtabular}
\begin{flushleft}
  $^{\rm a}$ Reference\cite{dzuba-11}. \\
  $^{\rm b}$ Reference\cite{safronova-09}. \\
  $^{\rm c}$ Reference\cite{porsev-12}- MBPT + corrections. \\
  $^{\rm d}$ Reference\cite{porsev-12}- All-order. \\
  $^{\rm e}$ Reference\cite{sahoo-11}. \\
  $^{\rm f}$ Reference\cite{olmschenk-07,olmschenk-09}. \\
  $^{\rm g}$ Reference\cite{pininngton-97}.
\end{flushleft}
\label{e1_tab}
\end{center} 
\end{table} 

The E1 transition amplitudes are presented in the Table. \ref{e1_tab}.
For comparison the results from the other theoretical and experimental works 
are also listed. Like HFS constant, transition amplitudes are calculated 
using the RCC wave functions. We have used the similar expressions and 
diagrams as the HFS except for one key difference, the hyperfine operator is 
replaced by the dipole operator. The experimental results are available only
for the $6s\;^2S_{1/2}-6p\;^2P_{1/2}$, $6s\;^2S_{1/2}-6p\;^2P_{3/2}$ and 
$6p\;^2P_{1/2}-5d\;^2D_{3/2}$ transitions. Among all the theoretical results, 
the results from the recent all-order work \cite{porsev-12} are closest to the 
experimental data. All other results, including ours, are on the higher side 
of the experimental value. The component wise contributions are listed in the 
Table. \ref{hfscompo_tab}. Like in the case of HFS constant, the DF term has
the dominant contribution. It contributes approximately about 118\%, 116\%, 
123\%, 124\% and 121\%, respectively, for the transitions listed in the table. 
A close agreement is observed in the DF data from our calculation with
the Ref. \cite{porsev-12}. 
%
%
\begin{table*}[ht]
\caption{Magnetic dipole hyperfine constants and E1 transition amplitudes, 
         contributions from different terms in the RCC. The operator ``O''
         here represents the hyperfine interaction Hamiltonian $H_{\rm hfs}$
         for HFS constant and the dipole operator $D$ for the E1 transition
         amplitudes.}
\label{tab-hfs-comp}
\begin{ruledtabular}
\begin{tabular}{cccccccccc}
State/Transition &\multicolumn{6}{c}{Coupled-cluster terms}            \\
\hline                                                           \\
      & DF & $\tilde O$ - DF & $S^\dagger_1\tilde O$ 
      & $S^\dagger_2\tilde O$
      & $S^\dagger_2\tilde O S_1$
      & $S^\dagger_1\tilde O S_1$
      & $S^\dagger_2\tilde O S_2$ & Other & Norm          \\
      &    &               & $+ c.c$ & $+ c.c$& $+ c.c.$ & & & &       \\
 \hline                                                          \\
$6s_{1/2}$& $9716.682$ & $-427.318$ & $2756.071$ & $1145.559$ & $125.344$
          & $204.287$ & $243.111$ & $-49.371$ & $-225.956$    \\
$6p_{1/2}$& $1548.208$ & $-49.951$ & $528.950$ & $245.862$ & $28.191$
          & $46.824$ & $23.105$ & $22.266$ & $-45.402$    \\
$6p_{3/2}$& $182.531$ & $-5.430$ & $56.715$ & $53.520$ & $5.802$
          & $4.570$ & $18.932$ & $1.897$ & $-5.012$    \\
$5d_{3/2}$& $289.667$ & $7.017$ & $82.212$ & $4.875$ & $3.863$
          & $5.924$ & $30.457$ & $4.316$ & $-7.197$    \\
$5d_{5/2}$& $110.234$ & $4.187$ & $29.781$ & $-220.089$ & $-16.443$
          & $2.032$ & $19.468$ & $1.186$ & $1.076$     \\
\hline                                                          \\
$6p_{1/2}\longleftarrow6s_{1/2}$& $3.242$ & $0.001$ & $-0.175$ & $-0.311$ & $-0.010$
          & $0.019$ & $0.029$ & $0.006$ & $-0.052$    \\

$6p_{3/2}\longleftarrow6s_{1/2}$& $4.543$ & $0.004$ & $-0.254$ & $-0.378$ & $-0.012$
          & $0.023$ & $0.037$ & $0.006$ & $-0.067$    \\

$5d_{3/2}\longleftarrow6p_{1/2}$& $3.861$ & $0.005$ & $-0.437$ & $-0.287$ & $-0.003$
          & $0.032$ & $0.034$ & $-0.004$ & $-0.068$    \\

$5d_{3/2}\longleftarrow6p_{3/2}$& $1.697$  & $0.002$ & $-0.206$ & $-0.121$
          & $-0.000$ & $0.012$ & $0.013$ & $-0.001$ & $-0.027$    \\

$5d_{5/2}\longleftarrow6p_{3/2}$& $5.200$ & $0.010$ & $-0.566$ & $-0.326$ & $-0.001$
          & $0.034$ & $0.036$ & $0.005$ & $-0.079$  
\end{tabular}
\end{ruledtabular}
  \label{hfscompo_tab}
\end{table*}


\subsection{NSI-E1PNC}

For calculation of NSI-E1PNC, we use the expression Eq. (\ref{e1pnccc1v})
derived for NSD-E1PNC in the electronic space. However, the important 
difference in this case is, as mentioned earlier, the PRCC cluster operators
are rank zero operators. In terms of diagrams, the ones with dominant 
contributions are derived from Fig. \ref{e1pnc_cc_fig} with the NSD-perturbed 
operators replaced by the NSI-perturbed ones. In Table. \ref{nsi_comp_tab}, we 
list the contributions from the different terms in PRCC. Among all the terms, 
the largest contribution, about 117\% of the total value, is from $DS^{(1)}_1$. 
The reason for this, as evident from Table. \ref{nsi_orbital_tab}, is the large 
$H_{\rm PNC}$ mixing between $6s\;^2S_{1/2}$ and $np\;^2P_{1/2}$ orbitals. 
This large contribution from $DS^{(1)}_1$ is consistent with the pattern of 
correlation reported in Ref. \cite{sahoo-11}. The next leading order 
contributions are $DS^{(1)}_2 + {\rm H.c.}$ and ${T^{(1)}_1}^\dagger D$. The 
former involve one core and one virtual orbitals, and the later connects 
${T^{(1)}_1}^\dagger $ and $D$ through a core orbital. These contribute about 
-17\% and 15\%, respectively. The terms 
${S^{(0)}_2}^\dagger DS^{(0)}_1 +{\rm H.c.}$ and 
${S^{(0)}_1}^\dagger DS^{(0)}_1+{\rm c.c.}$
are third and fourth leading order terms, contributing about -7\% each.
The contribution from normalization is -2.8\%. Small but not insignificant
contributions of 2\% and -1.8\% are also observed from the terms 
${T^{(0)}_2}^\dagger DT^{(1)}_1+{\rm c.c.}$ and ${S^{(1)}_1}^\dagger D$,
respectively.   

To examine the correlation pattern more closely we pick the leading order
terms $DS^{(1)}_1$ and ${T^{(1)}_1}^\dagger D$, and for these we calculate 
the E1PNC contributions from various intermediate $np_{1/2}$ states. The 
dominant contributions from these are tabulated in
Table. \ref{nsi_orbital_tab}. The same analysis but at the DF level 
is presented in Table. \ref{nsi_orbital_df}. As we see in both the tables, 
dominant contribution is from the $6p\;^2P_{1/2}$ state, contributing about 
117\% of the total value. The reason for this is, large $H_{\rm PNC}$ 
induced mixing with the energetically closer  $6s\;^2S_{1/2}$ state. 
The PRCC value is about 42\% larger than the Dirac-Fock contribution. This 
can be attributed to the large amplitude of $S^{(1)}_1$, and hence to 
the correlation effects incorporated using PRCC. The next dominant 
contribution among the core orbital is $5p_{1/2}$. This contributes to 
${T^{(1)}_1}^\dagger D$ through the $H_{\rm PNC}$ perturbed $6s\;^2S_{1/2}$. 
In this case as well PRCC contribution is larger than the DF.

The total NSI-E1PNC result from our calculation is presented in
Table. \ref{e1pnc_tab}. We have also listed the DF contribution. The other 
two theoretical results are based on the calculations with 
correlation-potential-method \cite{dzuba-11} and RCCSD(T) \cite{sahoo-11}. The 
E1PNC result from these two works differ from each other substantially. The 
CCSD(T) result from Ref. \cite{sahoo-11} is about 26\% larger than 
Ref. \cite{dzuba-11}. Our DF value is marginally on the higher 
side of the value reported in Ref. \cite{sahoo-11}. However, the 
total result lies between Refs. \cite{dzuba-11} and \cite{sahoo-11},
but closer to the coupled-cluster result of Ref. \cite{sahoo-11}.
%
%
\begin{table*}
\begin{center}
\caption{The NSI and NSD E1PNC component wise contribution from various terms in 
        the PRCC. The NSI and NSD contributions are listed in the units of 
        $iea_0\times10^{-11}(-Q_W/N)$ and  $iea_0 \mu'_W \times10^{-12}$,
        respectively.}
\begin{ruledtabular}
\begin{tabular}{ccccccccccccccc}
\multicolumn{2}{c}{Transition}& $D S^{(1)}_1$ & ${S^{(1)}}^\dagger_1 D$ & $D T^{(1)}_1$ & 
${T^{(1)}}^\dagger_1 D$ & $D S^{(1)}_2$ & ${T^{(0)}}^\dagger_1 D T^{(1)}_1$ & 
${T^{(0)}}^\dagger_2 D T^{(1)}_1$ & ${S^{(0)}}^\dagger_1 D S^{(1)}_1$ & 
${S^{(0)}}^\dagger_2 D S^{(1)}_1$ & 
${T^{(1)}}^\dagger_1 D S^{(0)}_2$ & Other & Norm \\
 &  & & & & & $+$c.c.&$+$c.c.&$+$c.c.&$+$c.c.&$+$c.c.&$+$c.c.& & &           \\
\hline                                   \\
 & & & & & & & NSI-PNC & & & & & &           \\
\hline
& & $8.950$&$-0.139$&$-0.005$&$1.179$&$-1.278$&$-0.025$&$0.168$&$-0.511$ &
$-0.529$ & $0.028$ & $0.003$ & $-0.215$    \\
                                 \\
$F_w$ & $F_v$ & & & & & &  NSD-PNC & & & & & &  &           \\
\hline
1 & 0 & $6.689$&$-3.301$&$0.030$&$1.118$&$-0.683$&$0.001$&$-0.250$&$-0.239$ &
$-0.279$ & $-0.008$ &$-0.099$&$-0.083$    \\
1 & 1 & $1.431$&$-2.539$&$-0.002$&$0.221$&$-1.221$&$0.000$&$-0.061$&$0.026$ &
$0.038$ & $0.008$ &$-0.021$&$0.059$    \\
2 & 1 & $-3.590$&$0.952$&$-0.020$&$-0.608$&$-0.114$&$0.000$&$0.130$&$0.163$ &
$0.194$ & $0.009$ &$0.053$&$0.079$    \\
\end{tabular}
\end{ruledtabular}
\label{nsi_comp_tab}
\end{center}
\end{table*}
%
%
\begin{table}
\begin{center}
\caption{The NSI E1PNC dominant contribution from the intermediate odd
         parity states in the PRCC. The listed E1PNC values are in the units 
         of $iea_0\times10^{-11}(-Q_W/N)$.}
\begin{ruledtabular}
\begin{tabular}{cccccccc}
\multicolumn{4}{c}{$D S^{(1)}_1$} &\multicolumn{4}{c}{${T^{(1)}}^\dagger_1D$}\\ 
\hline
 D & $S^{(1)}_1$ & E1PNC & state & D & $ {T^{(1)}}^\dagger_1$ & E1PNC & state \\
\hline
$-3.861$&$100.787$&$8.918$ &$6p_{1/2}$&$0.003$ &$-1.184$&$0.0$   &$2p_{1/2}$ \\ 
$-0.217$&$-29.596$&$-0.147$&$7p_{1/2}$&$-0.010$&$2.753$ &$-0.001$ &$3p_{1/2}$  \\ 
$0.047$ &$16.932$ &$-0.018$&$8p_{1/2}$&$-0.008$&$7.623$ &$-0.002$ &$4p_{1/2}$  \\ 
$-0.009$&$-21.986$&$-0.005$&$9p_{1/2}$&$1.290$&$39.984$ &$1.182$&$5p_{1/2}$ \\ 
$0.106$ &$-29.329$&$0.071$ &$10p_{1/2}$& & & \\ 
$-0.161$&$23.995$ &$0.088$ &$11p_{1/2}$& & & \\ 
$-0.096$ & $15.017$ & $0.033$& $12p_{1/2}$ & & & \\ 
\end{tabular}
\end{ruledtabular}
\label{nsi_orbital_tab}
\end{center}
\end{table}
%
%
\begin{table*}
\begin{center}
\caption{The Dirac-Fock dominant contributions from the intermediate odd
         parity states. 
         The listed NSI-E1PNC and NSD-E1PNC values are in the units of 
         $iea_0\times10^{-11}(-Q_W/N)$ and $iea_0\times10^{-12} \mu'_W$,
         respectively.}
\begin{ruledtabular}
\begin{tabular}{cccccccc}
\multicolumn{4}{c}{$D H_{\rm PNC}$} &\multicolumn{4}{c}{$H_{\rm PNC} D$}\\ 
\hline
 D & $ H_{\rm PNC}$ & E1PNC & state & D & $ H_{\rm PNC} $ & E1PNC & state \\
\hline
                                                                            \\
   &             &       & NSI-PNC &         &                &             \\
\hline
$-3.861$&$71.073$&$6.288$ &$6p_{1/2}$&$0.003$ &$-1.178$&$0.0$   &$2p_{1/2}$ \\ 
$-0.217$&$-18.657$&$-0.092$&$7p_{1/2}$&$-0.010$&$2.657$&$-0.001$ &$3p_{1/2}$\\ 
$0.047$ &$10.547$ &$-0.011$&$8p_{1/2}$&$-0.008$&$6.738$&$-0.001$ &$4p_{1/2}$\\ 
$-0.009$&$-13.620$&$-0.003$&$9p_{1/2}$&$1.290$&$23.438$&$0.693$&$5p_{1/2}$  \\ 
                                                                            \\
   &             &       & NSD-PNC &         &                &             \\
\hline
$-3.861$&$-118.154$&$6.210$ &$6p_{1/2}$&$0.003$ &$1.957$&$0.0$   &$2p_{1/2}$\\ 
$-0.217$&$31.016$&$-0.092$&$7p_{1/2}$&$-0.010$&$-4.417$&$-0.001$ &$3p_{1/2}$\\ 
$0.047$&$-17.534$&$-0.011$&$8p_{1/2}$&$-0.008$&$-11.202$&$-0.001$&$4p_{1/2}$\\ 
$-0.009$&$22.642$&$-0.003$&$9p_{1/2}$&$1.290$&$-38.965$&$0.684$&$5p_{1/2}$ \\ 
\end{tabular}
\end{ruledtabular}
\label{nsi_orbital_df}
\end{center}
\end{table*}
%
%


\subsection{NSD-E1PNC}

 For the NSD-PNC, the dominant contributions from various PRCC terms in 
Eq. (\ref{e1pnccc1v}) are listed in Table. \ref{nsi_comp_tab}.  For hyperfine 
transitions $F_v=0\rightarrow F_w=1$ and
$F_v=1\rightarrow F_w=2$, like in NSI-PNC, $DS^{(1)}_1$ is the leading 
order term. It contributes about 231\% and -130\%, respectively. 
For transition $F_v = 1 \rightarrow F_w = 1$, however, 
${S^{(1)}}^\dagger_1D$ is the dominant term, contributing about -123\%
of the total value. The same trend is reported in Ref. \cite{dzuba-11}, 
where the contributions are about 252\%, -226\% and -157\%, respectively, for 
the $F_v=0\rightarrow F_w=1$, $F_v=1\rightarrow F_w=1$ and 
$F_v=1\rightarrow F_w=2$ transitions. The next leading order term, 
${S^{(1)}}^\dagger_1D$, contribute about -114\% and 35\%, respectively
to the $F_v=0\rightarrow F_w=1$ and $F_v=1\rightarrow F_w=2$ transitions.  
However, the second leading order term is $DS^{(1)}_1$ for the
$F_v=1\rightarrow F_w=1$ transition, it contributes about 69\%. The next two 
leading order terms are ${T^{(1)}}^\dagger_1D$ and $DS^{(1)}_2 + {\rm H.c.}$. 
The contributions from these terms, in the sequence listed in  
Table. \ref{nsi_comp_tab}, are 39\%, 11\% and -22\%, and -24\%, -59\% 
and -4\%, respectively. Non-negligible contributions are also observed from 
the terms ${T^{(0)}_2}^\dagger DT^{(0)}_1+{\rm c.c.}$, 
${S^{(0)}_1}^\dagger DS^{(0)}_1+{\rm c.c.}$
and ${S^{(0)}_2}^\dagger DS^{(0)}_1+{\rm c.c.}$. 

 Unlike NSI-PNC, in NSD-PNC $np_{3/2}$ states also contribute to the E1PNC 
matrix element.  In Table. \ref{nsd_orbital_tab}, we list dominant 
contributions from odd parity $np_{1/2}$ and $np_{3/2}$ states in the PRCC 
calculations for the $F_v=0\rightarrow F_w=1$ transition, and specifically
the contributions from $DS^{(1)}_1$, ${S^{(1)}}^\dagger_1D$ and 
${T^{(1)}}^\dagger_1D$. At DF level we present the contributions from 
the $np_{1/2}$ states in Table. \ref{nsi_orbital_df}. As we see in these 
tables, both at DF and PRCC levels, the dominant contribution is from the 
$6p\;^2P_{1/2}$ state. The total contribution from this in the PRCC 
calculations is about 150\%, which can be attributed to the  230\% and -79\% 
contributions from $DS^{(1)}_1$ and ${S^{(1)}}^\dagger_1D$, respectively. The 
large contribution through $DS^{(1)}_1$ is due
to the strong $H_{\rm PNC}$ mixing with $6s\;^2S_{1/2}$. However that is not 
the case with $5d\;^2D_{3/2}$, which contributes through 
${S^{(1)}}^\dagger_1D$. At DF level $6p\;^2P_{1/2}$ contributes only through
mixing with $6s\;^2S_{1/2}$, and contribution is about 214\%.
The state $6p\;^2P_{3/2}$ is the third most dominant contributing one,
contributing about -40\%, through $H_{\rm PNC}$ perturbed $5d\;^2D_{3/2}$.
The other higher energy orbitals have negligible contribution.

The NSD-E1PNC total results are given in Table. \ref{e1pnc_tab}. For
comparison we have listed the DF contributions. Our DF results $6.915$,
$1.632$ and $-3.643$ for the three hyperfine transitions listed in 
Table. \ref{e1pnc_tab} compare well with the results $6.90$,
$1.70$ and $-3.70$, reported in Ref. \cite{porsev-12}. Our total results, 
however, are on the higher side of
the random-phase approximation (RPA) based results from Ref. \cite{porsev-12} 
for all hyperfine transitions. The other theoretical NSD-PNC data available
for comparison is from Dzuba {\rm et al} \cite{dzuba-11} using the 
correlation-potential-method.  Our results for transitions 
$F_v=0\rightarrow F_w=1$ and $F_v=1\rightarrow F_w=2$ are in good agreement 
with their results. For transition $F_v=1\rightarrow F_w=1$, however, our 
result is higher than their value.

%
%
\begin{table*}
\begin{center}
\caption{The NSD E1PNC dominant contribution from the intermediate odd parity 
         states in the PRCC. The E1PNC values are in the units of 
         $iea_0\times10^{-12} \mu'_W$. The contributions listed from the core
         $nP_{1/2}$ and $nP_{3/2}$ orbitals are from the terms 
         ${T^{(1)}}^\dagger_1 D$ and $DT^{(1)}$, respectively.}
\begin{ruledtabular}
\begin{tabular}{ccccccccccc}
\multicolumn{3}{c}{$D S^{(1)}_1$}&\multicolumn{3}{c}{${S^{(1)}}^\dagger_1D$} 
& Orbital & \multicolumn{3}{c}{${T^{(1)}}^\dagger_1 D$/ $DT^{(1)}$} & Orbital \\
\hline
 D & $S^{(1)}_1$ & E1PNC & D & $ {S^{(1)}}^\dagger_1$ & E1PNC &  & 
 D & ${T^{(1)}}^\dagger_1$ & E1PNC &  \\
\hline
$-3.861$  & $-126.570$ & $6.653$ & $3.242$ & $78.106$  & $-2.298$ & $6p_{1/2}$ & 
$0.003$ & $1.965$  & $0.000$ & $2p_{1/2}$ \\ 
$-0.217$  & $37.839$ & $-0.112$ & $-093$ & $-9.111$  & $-0.008$ & $7p_{1/2}$ & 
$-0.010$ & $-4.550$  & $-0.001$ & $3p_{1/2}$ \\ 
$0.047$  & $-21.776$ & $-0.013$ & $0.011$ & $4.704$  & $-0.001$ & $8p_{1/2}$ & 
$-0.008$ & $-12.416$  & $-0.001$ & $4p_{1/2}$ \\ 
$-0.009$  & $28.408$ & $-0.004$ & $0.013$ & $-5.743$  & $0.001$ & $9p_{1/2}$ & 
$1.290$ & $-63.798$  & $1.120$ & $5p_{1/2}$ \\ 
$0.106$  & $38.534$ & $0.056$ & $0.075$ & $-6.647$  & $0.005$ & $10p_{1/2}$ & 
&   &  & \\ 
$-0.161$  & $-32.883$ & $0.072$ & $-0.081$ & $4.431$  & $0.003$ & $11p_{1/2}$ & 
&   &  & \\ 
$-0.096$  & $-22.179$ & $0.029$ & $-0.053$ & $1.876$  & $0.001$ & $12p_{1/2}$ & 
&   &  & \\ 
         \\
\hline
$1.697$  & $-8.244$ & $0.000$ & $-4.543$ & $30.184$  & $-0.984$ & $6p_{3/2}$ & 
$-0.001$ & $-0.001$  & $0.000$ & $2p_{3/2}$ \\ 
$0.024$  & $2.418$ & $0.000$ & $0.358$ & $-5.101$  & $-0.013$ & $7p_{3/2}$ & 
$0.006$ & $0.002$  & $0.000$ & $3p_{3/2}$ \\ 
$0.008$  & $-1.376$ & $0.000$ & $-0.140$ & $2.786$  & $-0.003$ & $8p_{3/2}$ & 
$0.046$ & $0.109$  & $0.000$ & $4p_{3/2}$ \\ 
$-0.028$  & $1.845$ & $0.000$ & $0.136$ & $-3.670$  & $-0.004$ & $9p_{3/2}$ & 
$0.749$ & $-3.964$  & $0.021$ & $5p_{3/2}$ \\ 
$0.075$  & $-2.226$ & $0.000$ & $-0.050$ & $4.445$  & $-0.002$ & $10p_{3/2}$ & 
&   &  & \\ 
$0.080$  & $-1.527$ & $0.000$ & $0.022$ & $3.370$  & $0.001$ & $11p_{3/2}$ & 
&   &  & \\ 
$0.043$  & $-0.791$ & $0.000$ & $0.035$ & $1.675$  & $0.000$ & $12p_{3/2}$ & 
&   &  & \\ 
\end{tabular}
\end{ruledtabular}
\label{nsd_orbital_tab}
\end{center}
\end{table*}
%
%
\begin{table*}                        
\begin{center}                           
\caption{The total NSI (in the unit $iea_0\times10^{-11}(-Q_W/N)$) and 
         NSD (in the unit $iea_0 \mu'_W \times10^{-12}$) E1PNC results 
         compared with the previous theoretical results.} 
\begin{ruledtabular}                                           
\begin{tabular}{cccc}                                        
Transition & \multicolumn{2}{c}{This work}  &  Other works \\
\hline                                                                        
           &  DF   &  PRCC   &  \\
\hline                                                                        
          NSI-PNC &   &   &    \\
\hline                                                                        
$\langle 5d_{3/2}|\leftarrow \langle 6s_{1/2}|$ 
                             & $7.002$ & $7.626$ &
                               $6.262(20)^{\rm a},8.470^{\rm b}$ \\
                                                                 \\
             NSD-PNC &   &   &   \\
\hline                                                                        
$\langle 5d_{3/2},F_w=1|\leftarrow \langle 6s_{1/2},F_v=0|$ 
                              & $6.915$ &$2.896$ & 
                                $3.1(1.9)^{\rm a},2.6^{\rm c}$ \\
$\langle 5d_{3/2},F_w=1|\leftarrow \langle 6s_{1/2},F_v=1|$ 
                              & $1.632$ &$-2.061$& 
                                $-1.3(4)^{\rm a},-1.5^{\rm c}$ \\
$\langle 5d_{3/2},F_w=2|\leftarrow \langle 6s_{1/2},F_v=1|$ 
                              & $-3.643$ &$-2.753$& 
                                $-2.6(1.3)^{\rm a},-2.2^{\rm c}$ 
\end{tabular}                                                                 
\end{ruledtabular}                                                            
\begin{tabbing}                                                               
  $^{\rm a}$ Reference\cite{dzuba-11}.
  $^{\rm b}$ Reference\cite{sahoo-11}.
  $^{\rm c}$ Reference\cite{porsev-12}.
\end{tabbing}                                                                 
\label{e1pnc_tab}                                                               
\end{center}                                                                  
\end{table*}


\subsection{Uncertainty estimates}

To calculate the uncertainty of our E1PNC results we resort to a analysis based
on the sum-over-state approach. In this method, the net uncertainty associated 
with an intermediate state is
\begin{equation} 
 \Delta = \delta E^{\rm exci} + \delta E1 + \delta H_{\rm PNC}, 
\end{equation}
where $\delta E^{\rm exci}$ and $\delta E1$ are the deviations of excitation 
energy and E1 matrix element form the experimental data. And these are 
calculated based on results presented in Tables. \ref{ee_tab} and
\ref{e1_tab}, respectively. For $\delta H_{\rm PNC}$, uncertainty 
associated with $H_{\rm PNC}$ matrix, we resort to the deviation of
$\sqrt{A_iA_f}$ from experimental data, where $A_i$ and $A_f$ represent
the magnetic dipole hyperfine constants of initial and final states of the
E1PNC transition.

As discussed earlier, dominant contribution to NSI-PNC is from the 
$6p\;^2P_{1/2}$ state, which contributes through $DS^{(1)}_1$ i.e. 
$H_{\rm PNC}$ matrix element 
$\langle 6p\;^2P_{1/2}|H_{\rm PNC}|6s\;^2S_{1/2}\rangle$, 
$E1$ matrix $\langle 5d\;^2D_{3/2}|D|6p\;^2P_{1/2}\rangle$ and energy 
denominator $E_{6s_{1/2}}-E_{6p_{1/2}}$. 
The uncertainty associated with $H_{\rm PNC}$ matrix element, we get using 
our RCC results for hyperfine constants, is 9.1\%. 
And, the relative uncertainty of $E1$ matrix element and the energy 
denominator are calculated as 5.7\% and 3.4\%, respectively. 
Combining these, the net uncertainty in NSI-PNC result is 18.2\%.

For NSD-PNC as well $6p\;^2P_{1/2}$ is the dominant contributing
state. However in this case unlike NSI-PNC, apart from $DS^{(1)}_1$, 
contribution through ${S^{(1)}}^\dagger_1D$ is not negligible. The matrix
elements involve in this case are 
$\langle 5d\;^2D_{3/2}|H_{\rm PNC}|6p\;^2P_{1/2}\rangle$
and $\langle 6p\;^2P_{1/2}|D|6s\;^2S_{1/2}\rangle$, and the energy denominator
is $E_{6p_{1/2}}-E_{5d_{3/2}}$. Using the similar analysis we get 4.52\%,
11.26\% and 2.41\%, respectively for $\delta H_{\rm PNC}$, $\delta E1$ 
and $\delta E$. Combining these, we get 18.19\% as the net uncertainty associated
with term ${S^{(1)}}^\dagger_1D$. The $rms$ relative uncertainty in 
NSD-PNC results are then 18.17\%.


\section{Conclusions}

In this work, we present NSI and NSD-PNC transition amplitudes for the
$[4f^{14}]6s^2\;S_{1/2} - [4f^{14}]5d^2\;D_{3/2}$ transition in
$^{171}$Yb$^+$ ion. To estimate the uncertainty of the the PNC results,
we also calculate excitation energies, hyperfine structure constants and
E1 transition amplitudes for some of the important low lying states, using
RCC theory. The E1PNC results are computed using PRCC theory, which is 
formulated based on RCC theory and it incorporates electron 
correlation effects arising from a class of diagrams to all order in the 
presence of PNC interaction as a perturbation.

Our results for excitation energies, hyperfine structure constants and E1 
transition amplitudes are in good agreement, in some cases better, with the
previous experimental data. Our NSI-PNC result lies between the 
results from two previous studies reported in Refs. \cite{dzuba-11} and 
\cite{sahoo-11}.  The NSD-PNC DF results from our work is in excellent 
agreement with the results reported in Ref. \cite{porsev-12} for all hyperfine
transitions. The total NSD-PNC result for the 
$F_v=0\rightarrow F_w=1$  hyperfine transition lies between the results of
Refs. \cite{dzuba-11} and \cite{porsev-12}. For the remaining 
two, our results are slightly on the higher side. The upper 
bound to the theoretical uncertainty associated with E1PNC results
is about 20\%.


\begin{acknowledgments}
The author wish thank S. Chattopadhyay for useful discussions. The results 
presented in the paper are based on computations using the HPC cluster at 
Physical Research Laboratory, Ahmedabad.
\end{acknowledgments}


\end{document}